\documentclass[manuscript]{aastex}

\slugcomment{Not to appear in Nonlearned J., 45.}

\shorttitle{Miras in Sextans dSph galaxy}
\shortauthors{Sakamoto et al.}

\begin{document}

\title{Discovery of Mira variable stars in the metal-poor Sextans dwarf spheroidal galaxy}

\author{Sakamoto, Tsuyoshi\altaffilmark{1}}
\email{sakamoto@spaceguard.or.jp}
\author{Matsunaga, Noriyuki\altaffilmark{2,3}}
\author{Hasegawa, Takashi\altaffilmark{4}}
\and
\author{Nakada, Yoshikazu\altaffilmark{2}}

\altaffiltext{1}{Japan Spaceguard Association, 1716-3, Ookura, Bisei, Ibara, Okayama~714-1411, Japan}
\altaffiltext{2}{Kiso Observatory, Institute of Astronomy, School of Science, The University of Tokyo, 10762-30, Mitake, Kiso-machi, Kiso-gun, Nagano~397-0101, Japan}
\altaffiltext{3}{Department of Astronomy, The University of Tokyo, 7-3-1 Hongo, Bunkyo-ku, Tokyo 113-0033, Japan}
\altaffiltext{4}{Gunma Astronomical Observatory, 6860-86, Nakayama, Takayama, Agatsuma, Gunma~377-0702, Japan}

\begin{abstract}
We report the discovery of two Mira variable stars (Miras)
toward the Sextans dwarf spheroidal (dSph) galaxy.
We performed optical long-term  monitoring observations for two red stars
in the Sextans dSph.
The light curves of both stars in the $I_{\rm c}$ band
show large-amplitude (3.7 and 0.9 mag)
and long-period ($326\pm 15$ and $122\pm 5$ days) variations,
suggesting that they are Miras.
We combine our own infrared data with previously published data
to estimate the mean infrared magnitudes.
The distances obtained from the period-luminosity relation of the
Miras ($75.3^{+12.8}_{-10.9}$ and $79.8^{+11.5}_{-9.9}$ kpc , respectively), together with the radial velocities available,
support memberships of the Sextans dSph ($90.0\pm 10.0$ kpc).
These are the first Miras found in a stellar system with a metallicity
as low as  ${\rm [Fe/H]\sim -1.9}$, than any other known system with Miras.

\end{abstract}


\section{Introduction}

Miras are pulsating stars
with initial masses between 0.8 and 8 solar masses
in the Asymptotic Giant Branch (AGB) phase,
and eject material via stellar winds
into the interstellar medium (e.g., Habing~1996).
The ejected material contain chemical elements 
that have been dredged up from the interior
(e.g., carbon and s-process elements).
A large amount of dust forms in the ejecta of the
Miras, and then such dust grains regulate
the cooling of the interstellar medium and the fragmentation
of collapsing molecular clouds into stars. 
Thus, Miras play an important role in
providing the heavy elements and dust grains
from the early Universe to the present day.

The lack of appropriate theoretical models of Miras
currently prevents us from determining their fundamental properties,
such as metallicities, based on observational data for Miras.
Therefore, Miras in stellar systems with known metallicity and/or age
distribution
can be important tracers to study the evolution of Miras and their impacts
on chemical enrichment. For example, most of the 
Galactic globular clusters are old stellar systems with 
a single metallicity  or 
a narrow metallicity distribution,
offering an important sample of low-mass and metal-poor Miras
(Frogel \& Whitelock 1998; Feast et al.~2002). 
We note that Miras are found only in
the clusters with ${\rm [Fe/H]>-1}$.
Another interesting sample is  low- to intermediate-mass Miras 
found in the Magellanic Clouds.
A formidable amount of literature exists on those objects
(e.g.~Ita et al. 2004ab; Fraser et al. 2008; Soszy\'{n}ski et al. 2009;
Groenewegen et al. 2009).

Galactic dSphs provide us with a sample of even lower metallicity
objects than  the globular clusters.
It is known that
the fainter galaxy tends to have the lower mean metallicity
(Norris et al.~2010). 
Therefore, the faint dSphs are excellent places to study metal-poor
Miras if any Mira is found.
Recent monitoring surveys in Galactic dSphs have discovered
several Miras 
(Fornax dSph, Whitelock et al. 2009; Leo~I dSph, Menzies et al.~2010;
Sagittarius dSph, Lagadec et al. 2009; Sculptor dSph, Menzies et al.~2011).
Among the dSphs with previously known Miras, the Sculptor dSph is 
the most metal deficient one, which has the metallicity distribution
with a peak at ${\rm [Fe/H]}=-1.56$ 
and a dispersion of 0.48 (Kirby et al. 2009). 
Sloan et al. (2009) reported the evidence of circumstellar dust around
one of the Miras in the Sculptor dSph (Menzies et al.~2011),
based on the {\it Spitzer Space Telescope} spectroscopy.
This suggests that AGB stars in low-metallicity
environments can make a significant contribution to dust formation
in the early Universe.

Our target galaxy, the Sextans dSph, shows 
a metallicity distribution with a peak at 
${\rm [Fe/H]}=-1.9$
and is one of the most metal-poor dSphs in the Galaxy
(Battaglia et al.~2011). So far, two monitoring surveys 
have been conducted for the center of this galaxy,
revealing dozens of short-period variable stars (Mateo et al. 1995;
Lee et al. 2003).
No Mira has been found previously.
In this paper, we report the discovery of two Miras toward
the Sextans dSph. In Section 2, we describe optical and infrared photometric observations, 
and discuss their membership to the Sextans dSph. In Section 3, we
discuss the chemical properties and their impacts.

\section{Observations and Results}

\subsection{Our targets}

In order to explore Miras in the Sextans dSph,
we selected two target stars from the photometric catalogs presently available.
These are listed in Table~\ref{tab:Object}.
The first target \#1, SDSS~J101525.93$-$020431.8, was selected
using the color criteria, 
$J-H>0.7$, $H-K_{\rm s}>0.3$ on the 2MASS catalog (Skrutskie et al.~2006)
and $g-r>0.8$, $r-i>0.3$ on the SDSS catalog (Adelman-McCarthy et
al.~2008).
These criteria are also used 
for our monitoring survey of Miras in the Galactic halo
(Sakamoto et al., in preparation). 
The target \#1 is carbon-rich, showing a spectrum
with the strong CN absorption band at 7900${\rm \AA}$ 
(Mauron et al. 2004; Cruz et al. 2007). 
The second target \#2, SDSS~J101234.29$-$013440.8, was later added
because of its variation detected in QUEST1 
(QUasar Equatorial Survey Team, Phase 1) variability survey 
(Rengstorf et al. 2009).
The $R$-band light curve of \#2 over 2 years 
in the QUEST1 showed a variation with a large amplitude 
($\Delta R \geq 1.2$ mag) and a long period (over 100 days),
although their time sampling was
not good enough to estimate the period.
The target \#2 is oxygen-rich, showing a spectrum
with clear TiO molecular absorption lines (Suntzeff et al. 1993).

\subsection{$I_c$-band photometry}

We conducted photometric monitoring observations of the two selected targets
in the direction of the Sextans using the 2KCCD camera attached
to the 105-cm f/3.0 Schmidt telescope at Kiso Observatory (Itoh et al.~2001).

The observations started in December 2008 and February 2010
for the targets \#1 and \#2, respectively, and were repeated until February 2012.
Time series $I_c$-band images were obtained.
The data were reduced following standard procedures with IRAF,
including bias subtraction (both the level of the overscan region in each image
and the bias pattern taken on each night) 
and the flat-field correction with $I_c$-band dome-flat images.
Instrumental magnitudes of the targets and comparison stars
were measured with aperture
photometry using the IRAF/APPHOT package. 
The comparison stars were selected from the SDSS database 
(Adelman-McCarthy et al. 2008), and
their $I_c$ magnitudes were 
calculated by using the transformation of 
Jordi et al. (2006),
\begin{equation}
I_c=i'+(-0.386\pm 0.004)(i'-z')-(0.397\pm 0.001).
\end{equation}
Using these $I_c$ magnitudes for calibrating the magnitude scale,
we obtained the magnitudes of our target stars as listed in Tables 2 and 3.

\begin{figure}
\caption{
(Top): $I_c$-band light curves of our targets.
The arrows indicate the epochs of the IRSF near-infrared data.
(Bottom): Phased light curves.
For \#1 the long-term trend is substracted. 
\label{fig:LC}
}
\epsscale{1.0}
\plottwo{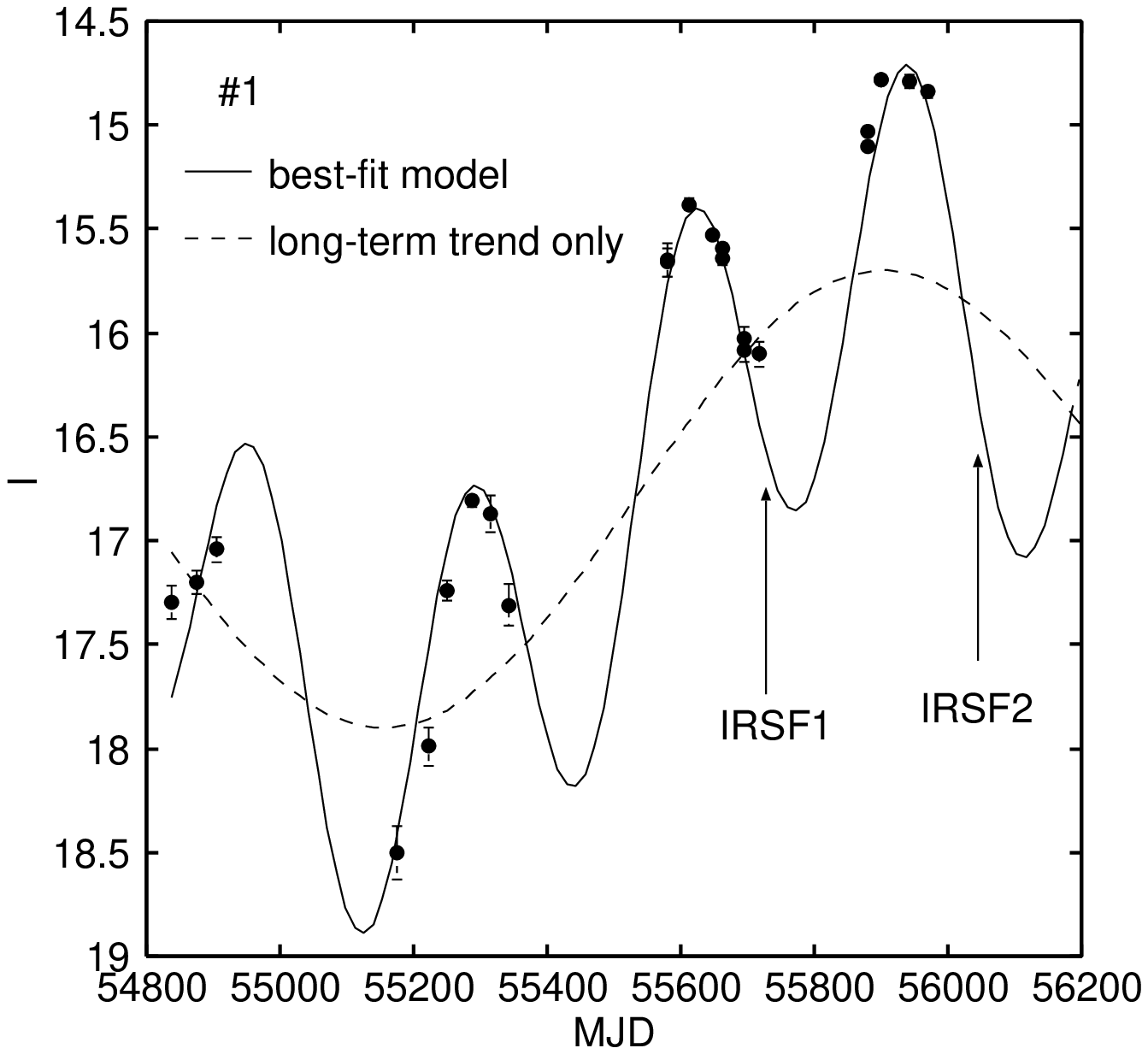}{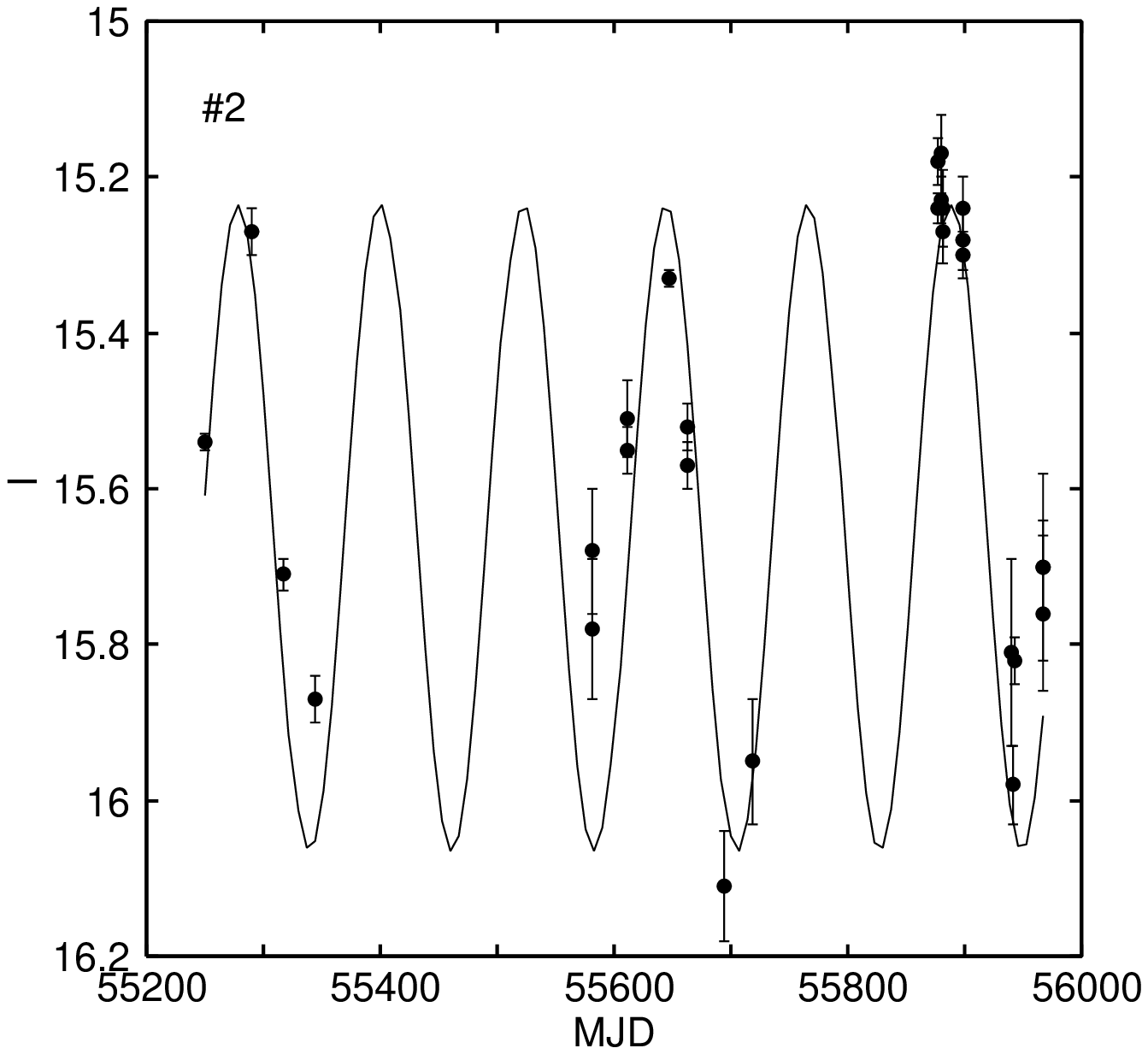}
\plottwo{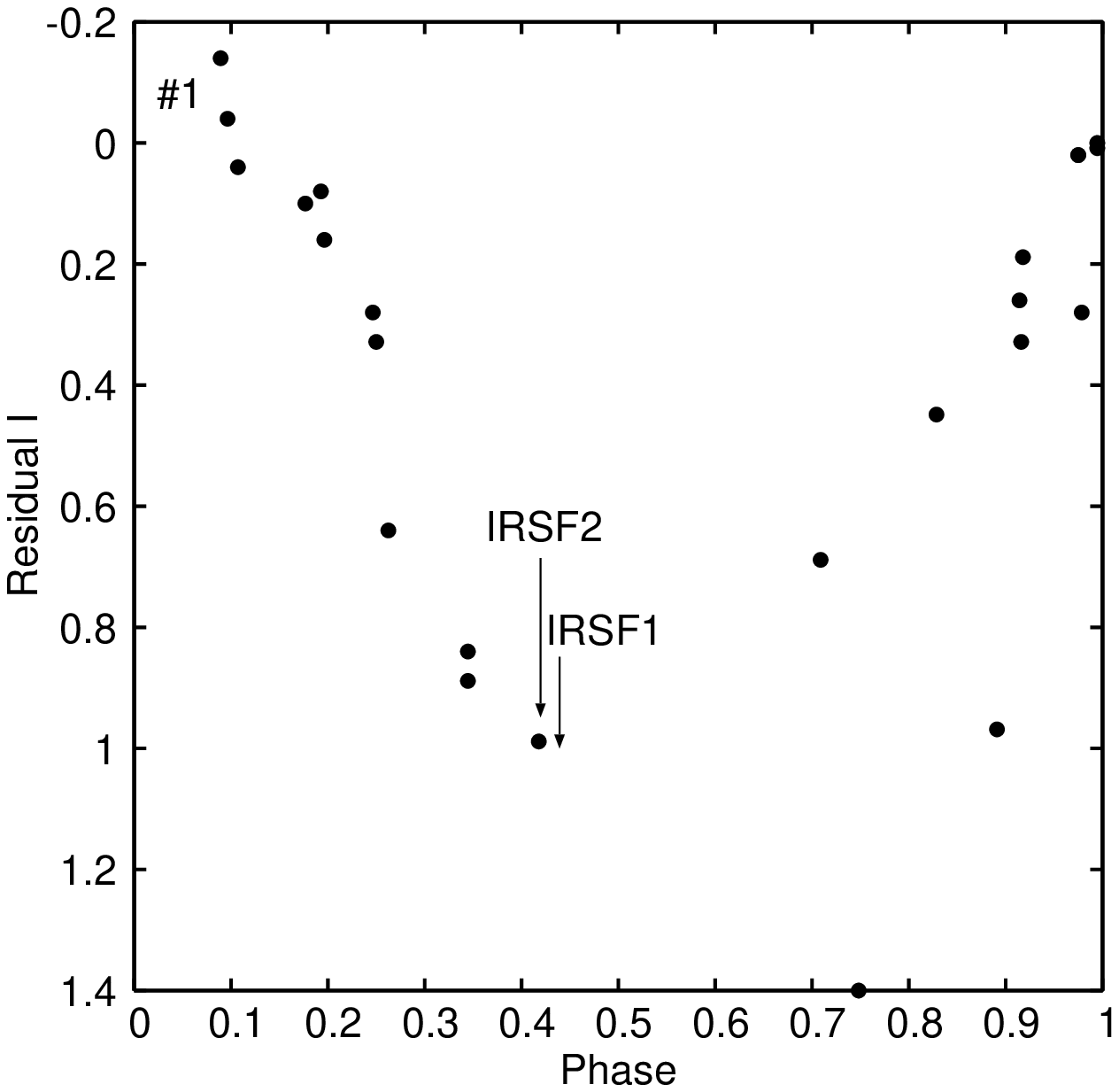}{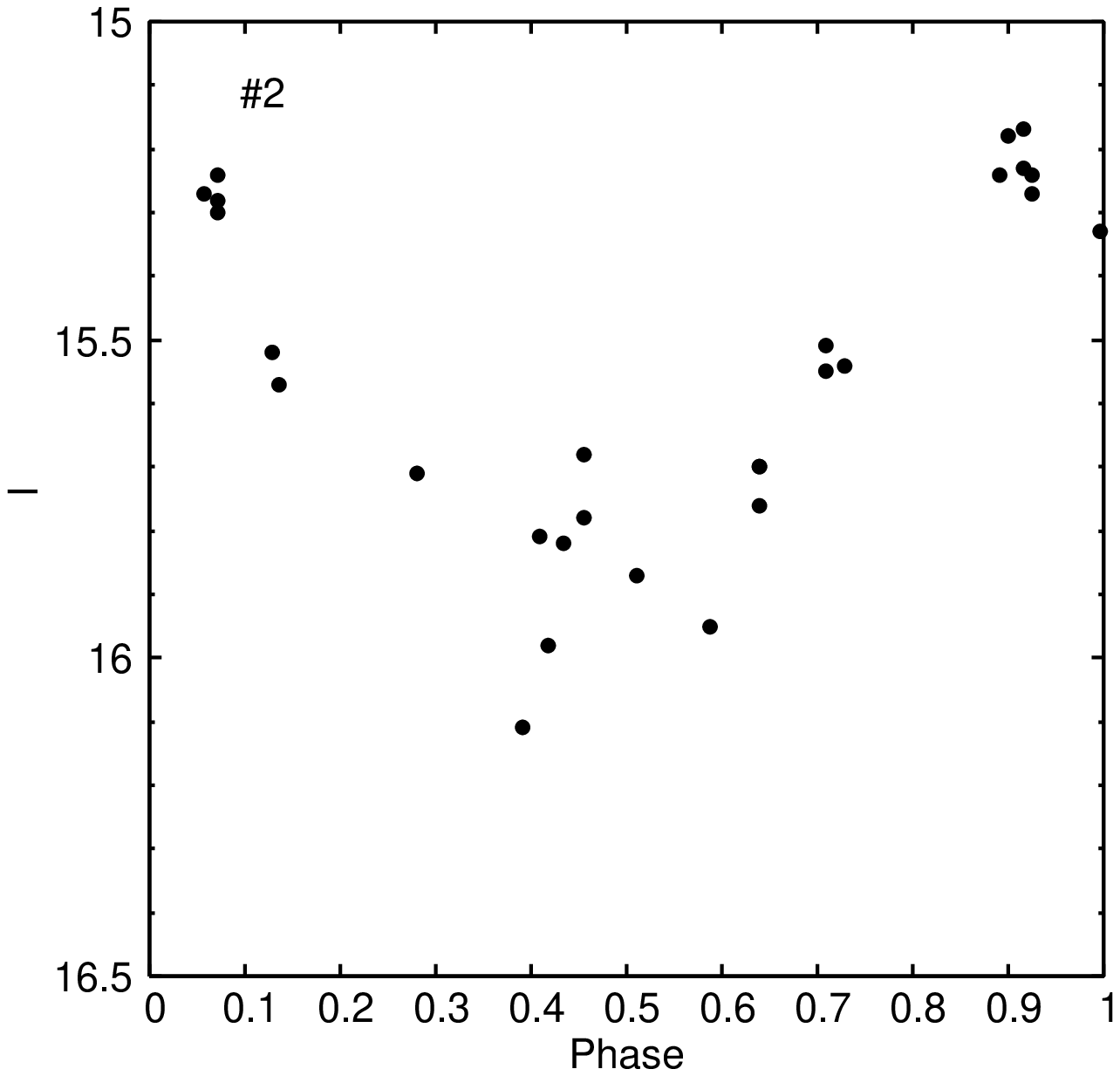}
\end{figure}

Fig.~\ref{fig:LC} plots the $I_c$ variations
against the Modified Julian Date (MJD). Both stars
show the long-period and large-amplitude variation characteristic
of either Miras or semi-regular variables.
The peak-to-valley amplitudes ($\Delta I_c$) are
3.72 mag for the target \#1 and 0.94 mag for the target \#2.
Miras are generally considered to have the $I_c$ amplitude greater 
than 0.9-1.0 mag
(Ita et al. 2004ab; Matsunaga et al. 2005).
The target \#1 is clearly a Mira, whereas the target \#2 falls between Miras
and semi-regulars.

The light curve of the target \#1 shows a clear modulation
over the entire observation run
indicating a long-term variation
which is often observed for carbon-rich Miras 
(Whitelock et al. 2003). We subtract this long-term trend
which is fitted by a sine curve with a period of 1500~days.
The residual light curve shows a regular periodic variation as expected.
Then, we applied the Phase Dispersion Minimization (PDM, Stellingwerf 1978)
to the residual curve to obtain a period 326~days as the best estimate.
It should be noted that due to the insufficient coverage of the light curve,
especially around the expected minima, we cannot exclude the possibility
of the star having a period roughly half of the 326~days based only on
our $I_c$ band data. Nevertheless, the shorter period is unlikely,
considering that the star is so red, $(H-K_{\rm s}) \gtrsim 1.0$, 
and  all of such red LMC Miras in Ita et al. (2004) 
have periods of 300$-$500 days.
We examined various assumptions on the long-term trend
(e.g., sine curves with different periods and/or amplitudes),
and the estimated period stays within 15 days from the above value.
Thus, the period of \#1 is estimated to be $326\pm 15$ days. 
In contrast, the light curve of the target \#2 appears to show
no long-term trend, and the period of 122 days is obtained with the PDM.
Sine curves with periods of 117$-$127 days 
seems consistent with the photometric data.
Thus, the periods of \#2 is estimated to be  $122\pm 5$ days.

\subsection{Near-infrared photometry}

For the target \#1, we obtained near infrared images in the
$J$, $H$, and $K_{\rm s}$-bands using
the IRSF 1.4-m telescope at South African Astronomical Observatory
(Nagayama et al. 2003).
Our near-infrared observations were carried out on June 14, 2011
and April 28, 2012,
and the magnitudes of the target were calibrated based on
2MASS magnitudes of the neighboring stars.
In addition, the near-infrared magnitudes for our targets were collected
from a few large-scale catalogues (2MASS, Skrutskie et al. 2006;
UKIRT DR6, Lawrence et al. 2007; DENIS, Epchtein et al. 1994).
Table~\ref{tab:Kband} lists the magnitudes available from these sources,
where the magnitudes were transformed into those in the IRSF photometric
system (Kato et al. 2007).
The magnitudes in the DENIS and  UKIRT DR6  photometric systems
were transformed into the 2MASS system 
by using the equations in Carpenter (2001) and Hewett et al. (2006),
respectively,
and then further transformed into
the IRSF system by using Kato et al. (2007).
The $J$, $H$, and $K_{\rm s}$ magnitudes list in Table~\ref{tab:Kband}
and show significant variations.
The mean magnitudes 
are obtained by taking averages of four or three independent data in Table 4.

Here we check the uncertainty in the transformation 
between the different catalogs
by comparing the magnitudes of stars around our Miras after the transformation 
into the IRSF magnitude.
The infrared magnitudes well match between the 2MASS, UKIRT, and 
IRSF photometry measurements within their uncertainties, whereas
the $K_{\rm s}$ magnitudes obtained from the DENIS are 
 different from  those in the others by $\sim 0.3$.
We added a shift to the above DENIS magnitudes
to fit them onto the 2MASS magnitude scale
(the magnitudes in Table 4 are on this corrected scale).

We estimate the uncertainties in the random-phased mean magnitudes in the 
$K_{\rm s}$-band as follows. 
Carbon-rich Miras with periods of 300$-$350 days in
the solar neighborhood and nearby dwarf galaxies
appear to have $K_{\rm s}$-band amplitudes of 0.5--1.0 mag
(Whitelock et al. 2003).
O-rich Miras with periods of 100$-$150 days have the $K_{\rm s}$-band
amplitude of 0.14$-$0.88 mag in the sample of Whitelock et al. (2000).
The amplitudes of \#1 and \#2 are assumed to be 0.75 mag and
0.51 mag in the $K_s$-band, respectively,
suggesting that the observed mean magnitudes are
located within $\pm$ 0.38 mag and $\pm$ 0.26 mag around 
their true mean magnitudes, respectively.
Our infrared datasets consist of four and three independent data for \#1 and \#2, 
and thus the uncertainties in the mean magnitudes are approximately
0.19 mag and 0.15 mag.

\subsection{Distances and radial velocities}

To discuss whether or not our target stars are associated 
with the Sextans dSph, we estimate the distances on the basis of the
period-luminosity (PL) relation of Miras.
Miras, as well as semi-regulars with a relatively regular variation
and an amplitude close to 1 mag, like \#2,
are generally found to lie on the Mira PL relation.
Thus, we can apply this relation to both of our targets (Ita et al. 2004ab). 

Adopting the distance modulus of the LMC to be $18.50\pm 0.02$ (Alves 2004),
the  PL relation for the Miras (Ita \& Matsunaga 2011) is expressed as
\begin{equation}
  M_K=(-3.675\pm 0.076)\log P+(1.456\pm 0.173).
\end{equation}
Ita \& Matsunaga (2011) also found that 
the redder Miras tend to show large offsets from the PL relation.
They suggested that such offsets are caused by the dust
shells around red Miras ($J-K_{\rm s}\gtrsim 2$),
and we discuss the implication for our Miras in Section 3.
For \#1, the offset in the $K_{\rm s}$ band is estimated to be 
$0.12\pm 0.09$ mag (Ita \& Matsunaga 2011).
In contrast, the target \#2 is so blue that such an offset is negligible,
if any.
The interstellar extinctions in the Galaxy
in the $K_{\rm s}$ band
are calculated to be $0.02$~mag from the map in Schlegel et al. (1998). 

After correcting the above offset and the interstellar extinction,
we obtain distances of $75.3^{+12.8}_{-10.9}$ kpc
and $79.8^{+11.5}_{-9.9}$ kpc for the targets \#1 and \#2.
The uncertainties in the distances include those in periods, 
period-luminosity relation, and mean $K_{\rm s}$-band magnitudes.
The estimated distances are consistent with
that of the Sextans dSph, $90.0\pm 10.0$~kpc (Lee et al. 2003, 2009).

Here we discuss the effect of the long-term trend on the distance
esitmate for \#1.
Long-term trends have been detected from optical and infrared wavelengths,
however the relationship among different wavelengths has not been 
established, in particular when our data points are poor.
In addition, the observation dates for the 2MASS and the UKIRT 
were MJD 51174 and 53726, respectively, 
whereas our $I_c$ observations started on MJD=54836.
The long interval between the infrared and the $I_c$ observations makes the
estimation of the long-term trend difficult.
However, the long-term trend is suggested to be partly related with 
absorption by the circumstellar dust (Winters et al. 1994).
The circumstellar extinction is also suggested to be 
correlated with the deviation from the PL relation (Ita \& Matsunaga 2011),
and its deviation is corrected using the ($H-K_{\rm s}$) color in this work.
Thus, the above correction considering the ($H-K_{\rm s}$) color works,
at least partly, as the correction of the long-term trend.
In fact, the UKIRT photometry indicates that 
the \#1 was both fainter and redder at its epoch
than at the epochs of other photometry.      
Even if we exclude the UKIRT magnitudes from our calculation,
with the proper correction applied,      
we get the same result within the uncertainty.
 

We check the reliablity of the distance for \#2.
The Ic-band PL relation in Ita \& Matsunaga (2011) is relatively tight
near the period of our Mira \#2 (P=122 days),
although it has a higher dispersion than the Ks-band PL relation.
The distance based on the Ic-band PL relation 
is estimated to be $86.6^{+23.1}_{-18.2}$~ kpc,
consistent with the distance by the Ks-band PL relation 
($79.8^{+11.5}_{-9.9}$~kpc).

The systematic radial velocity of the Sextans dSph is 
226.0 ${\rm km~s^{-1}}$
with a dispersion of 8.4~${\rm km~s^{-1}}$ (Battaglia et al. 2011).
For the targets \#1 and \#2, the radial velocities at a single epoch are
$202\pm 12$~km~s$^{-1}$ (Mauron et al. 2004)
and $228.2\pm 2$~km~s$^{-1}$ (Suntzeff et al. 1993).
The radial velocity of a Mira 
measured with the optical spectra shows a time variation
due to the pulsation by 10--25 ${\rm km~s^{-1}}$ (Alvarez et al. 2001).
The radial velocities of foreground stars
in the direction of the Sextans dSph
show a peak around 0--50 ${\rm km~s^{-1}}$ and 
a weak tail toward a large velocity.
Probabilities of foreground stars with the velocities of our targets are low,
and this clearly support the memberships to the Sextans dSph.

Another possible system that the targets might be associated with is
a stellar stream in the Galactic halo. The SDSS reveals 
the presence of a 60-degrees-long stream
extending from the Ursa Major to the Sextans (e.g., Grillmair 2006).
However, its heliocentric distance is about 20 kpc, 
much smaller than those of our targets.
The model of Law et al. (2005) predicts
the presence of the Sagittarius stream at 
a heliocentric distance of 20-60 kpc 
toward the Sextans dSph in the oblate dark-matter halo of the Galaxy,
although the stream has not been identified in its direction.
Furthermore, the predicted radial velocities are 250--290 km~s$^{-1}$,
which are larger than the measured values of the targets.
Thus, both of  our targets are most likely associated 
with the Sextans dSph.

\section{Discussion}
As mentioned in Section 1, our targets in the Sextans dSph are
expected to be metal-poor.
Further constraints can be inferred from their locations in the galaxy.
The target \#1 is located within the tidal elliptical radius of 
the Sextans dSph (160 arcmin, ellipticity of 0.35),  
but in its outer part.
Unfortunately, the metallicities for a large sample of stars around the
target \#1 are not yet available.
Nevertheless, Battaglia et al. (2011) reported 
that the Sextans dSph has a clear metallicity gradient:
most stars beyond an elliptical radius of 48 arcmin (ellipticity 0.35)
in projection
appear to have [Fe/H]$<-$2.2, while metal-rich stars
are concentrated toward the center.
The target \#1 is located beyond the elliptical radius of 48 arcmin, 
suggesting that its metallicity is very low.

As shown in Section 2.4, the target \#1 is very red, 
in particular at the UKIRT epoch, $J-K_{\rm s}=2.96$.
Ita \& Matsunaga (2011) suggested the existence of
the dust shells around such red Miras from their magnitudes 
fainter than the PL relation.
The deviation of the $K_{\rm s}$ magnitude in each catalog
from that of the period-luminosity relation
appears to follow a function of the ($J-K_{\rm s}$) and ($H-K_{\rm s}$)
in Ita \& Matsunaga (2011).
Sloan et al. (2009, 2012) and Matsuura et al. (2007) detected 
the SiC dust excess for the spectra of the carbon-rich AGB stars 
in nearby metal-poor dSphs.
The color and period of target \#1 are similar to those in
the carbon-rich Miras in Sloan et al. (2012).
Thus, we suggest a possible presence of the circumstellar dust
around carbon-rich Miras with the metallicity $\sim$ 100 times
lower than the solar.
The Sextans dSph has a lower mean metallicity than 
any other dSph in Sloan et al. (2012),
and the target \#1 can impose an important limit on the metallicity dependence
of the dust content around carbon-rich Miras. 

The target \#2 is located in the inner part of the Sextans dSph.
In contrast to the outer part, 
relatively metal-rich stars ([Fe/H]$>-1.0$) coexist 
with the metal-poor stars at the inner part,
and thus the limit on the metallicity of \#2 is not so strong.
The short-period and oxygen-rich chemistry suggest
that the target \#2 is similar to
low-mass Miras found in Galactic globular clusters.

\section{Summary}

We discovered two Miras ($P=326$ and 122~days, respectively)
toward the metal-poor Sextans dSph
by performing photometric monitoring 
for red stars over periods of 3-4 years, 
although the shorter-period one may be a semi-regular.
The distances and radial velocities of the objects are consistent
with those of the Sextans dSph, which suggests that
the two objects belong to the Sextans dSph.
Thus, these objects are found in the lowest metallicity dSph with Miras, 
[Fe/H]$<-2$.
The red near-infrared color of the longer-period Mira \#1 suggests 
the presence of circumstellar dust in a system with the lowest metallicity.
Follow-up spectroscopic observations in the infrared wavelengths 
is useful to investigate dust formation 
in the very metal-poor objects.

\acknowledgments

We thank Aoki, T., Soyano, T., Tarusawa, K., and Dr. Mito, H. for supporting our observations at Kiso observatory.      
We are grateful to Dr. Feast, M. W. for reading the manuscript carefully and for his comments.
TS thanks Dr. Ita, Y. for providing the source code of the period determination.
NM acknowledges the support of Grant-in-Aid for Young Scientists
(No~23684005) from the Japan Society for the Promotion of Science (JSPS).
Funding for the SDSS and SDSS-II has been provided by the Alfred P. Sloan Foundation, the Participating Institutions, the National Science Foundation, the U.S. Department of Energy, the National Aeronautics and Space Administration, the Japanese Monbukagakusho, the Max Planck Society, and the Higher Education Funding Council for England. The SDSS Web Site is http://www.sdss.org/.

\clearpage
\clearpage
\rotate
\begin{deluxetable}{llllllllll}
\tablecaption{List of our target stars.
The magnitudes $<I_c>$ denote the mean magnitude of the targets
in the $I_c$-band, whereas $\Delta I_c$ show their amplitudes.
The distances obtained by us are listed as well as the 
radial velocities and the chemistry types.
\label{tab:Object}
}
\tablewidth{0pt}
\tablehead{
\colhead{ID}
&\colhead{RA}
&\colhead{DEC}
&\colhead{Period}
&\colhead{$<I_c>$}
&\colhead{$\Delta I_c$}
&\colhead{Distance}
&\colhead{Radial velocity}
&\colhead{Carbon-rich/}\\
\colhead{}
&\colhead{(J2000)}
&\colhead{(J2000)}
&\colhead{(days)}
&\colhead{(mag)}
&\colhead{(mag)}
&\colhead{(kpc)}
&\colhead{(${\rm km~s^{-1}}$)}
&\colhead{Oxygen-rich}
}
\startdata
\#1 (SDSS~J101525.93$-$020431.8)
&10:15:25.93&$-$02:04:31.8&$326\pm 15$
&17.0&3.72
&$75.3^{+12.8}_{-10.9}$
&$202\pm 12$\tablenotemark{c}
&Carbon-rich\tablenotemark{c}\tablenotemark{d}\\
\#2 (SDSS~J101234.29$-$013440.8)&10:12:34.29&$-$01:34:40.8&$122\pm 5$
&15.7&0.94
&$79.8^{+11.5}_{-9.9}$
&$228.2\pm 2.0$\tablenotemark{e}&Oxygen-rich\tablenotemark{e}\\
Sextans dSph &10:13:03&$-$01:36.9&---&---&---&$90.0^{+10.0}_{-10.0}$\tablenotemark{a}&$226\pm 8.4$\tablenotemark{b}&---\\
\enddata
\tablenotetext{a}{Average of Lee et al. (2003) and Lee et al. (2009)}
\tablenotetext{b}{Battaglia et al. (2011)}
\tablenotetext{c}{Mauron et al. (2004)}
\tablenotetext{d}{Cruz et al. (2007)}
\tablenotetext{e}{Suntzeff et al. (1993)}
\end{deluxetable}

\clearpage

\begin{table}
\caption{Time-series photometric results for our target \#1.
Modified Julian Date (MJD), the $I_c$-band magnitude, and the
photometric error is listed for each image we took.\label{tab:Iband1}}
\small
\begin{tabular}{cccccc}
\hline
\multicolumn{6}{l}{\#1~(SDSS~J101525.93$-$020431.8)}\\
MJD & $I_c$~(mag) & $e_{I_c}$~(mag) &MJD & $I_c$~(mag) & $e_{I_c}$~(mag)\\ 
\hline
54836.77&17.30&0.08&55646.49&15.53&0.01\\
54875.69&17.20&0.06&55662.53&15.59&0.02\\
54904.57&17.04&0.06&55663.49&15.64&0.03\\
55174.78&18.50&0.13&55694.50&16.08&0.06\\
55221.64&17.99&0.09&55694.50&16.03&0.06\\
55249.61&17.24&0.05&55718.48&16.10&0.06\\
55288.48&16.81&0.03&55879.84&15.03&0.02\\
55314.58&16.87&0.09&55880.82&15.10&0.02\\
55342.49&17.31&0.10&55899.85&14.78&0.01\\
55580.79&15.66&0.07&55899.85&14.78&0.02\\
55580.79&15.65&0.08&55942.73&14.79&0.03\\
55611.56&15.38&0.03&55970.55&14.84&0.03\\
\hline
\end{tabular}
\end{table}

\begin{table}
\caption{Same as Table 2, but the target \#2.\label{tab:Iband2}}
\small
\begin{tabular}{ccccccccc}
\hline
\multicolumn{9}{l}{\#2~(SDSS~J101234.29$-$013440.8)}\\
MJD & $I_c$~(mag) & $e_{I_c}$~(mag) &MJD & $I_c$~(mag) & $e_{I_c}$~(mag) & MJD & $I_c$~(mag) & $e_{I_c}$~(mag) \\
\hline
55249.60&15.54&0.01&55663.49&15.57&0.03&55898.73&15.24&0.04\\
55289.42&15.27&0.03&55694.50&16.11&0.07&55898.73&15.30&0.03\\
55316.47&15.71&0.02&55718.47&15.95&0.08&55939.77&15.81&0.12\\
55344.47&15.87&0.03&55876.77&15.24&0.02&55940.76&15.98&0.05\\
55580.78&15.68&0.08&55877.78&15.18&0.03&55942.72&15.82&0.03\\
55580.78&15.78&0.09&55879.83&15.17&0.05&55967.70&15.70&0.12\\
55611.55&15.55&0.03&55879.84&15.23&0.03&55967.70&15.76&0.10\\
55611.55&15.51&0.05&55880.81&15.24&0.05&55967.70&15.70&0.06\\
55646.65&15.33&0.01&55880.81&15.27&0.04&&&\\
55662.53&15.52&0.03&55898.72&15.28&0.04&&&\\
\hline
\end{tabular}
\end{table}
\begin{table}
\caption{$J$, $H$, and $K_{\rm s}$ magnitudes of our targets
available in our photometry with the IRSF/SIRIUS 
and other near-infrared catalogs.The magnitudes in other photometric systems were transformed into the IRSF system.
\label{tab:Kband}}
\small
\begin{tabular}{ccccccccc}
\hline
Object& \multicolumn{4}{c}{\#1} & \multicolumn{4}{c}{\#2} \\
Filter&MJD      & $J$     & $H$     & $K_{\rm s}$ &MJD& $J$     & $H$     & $K_{\rm s}$\\
\hline      
2MASS&51174.35 &$14.017\pm 0.026$& $12.908\pm 0.025$& $11.981\pm 0.026$&
51174.31&$14.128\pm 0.022$& $13.567\pm 0.029$& $13.391\pm 0.036$\\  
UKIRT&53726.68&$15.142\pm 0.005$& $13.596\pm 0.004$& $12.179\pm 0.004$&
53726.68&$13.853\pm 0.002$& $13.303\pm 0.002$& $13.139\pm 0.003$\\
IRSF1&55726.76& $13.830\pm 0.005$& $12.470\pm 0.005$& $11.430\pm 0.005$&
---&---&---&---\\
IRSF2&56045.76&$13.420\pm 0.005$& $12.470\pm 0.005$& $11.430\pm 0.005$&---&---&---&---\\
DENIS&---&---&---&---&50480.25&---&---&$13.34\pm 0.17$\\
Average&---&14.10& 12.86& 11.76&---&13.99& 13.43&13.34\\
\hline
\end{tabular}
\end{table}
\end{document}